# Fieldable muon spectrometer using multi-layer pressurized gas Cherenkov radiators and its applications

Junghyun Bae[1,*], Stylianos Chatzidakis[1]

[1]School of Nuclear Engineering, Purdue University, West Lafayette, IN 47906

**Abstract**

Cosmic ray muons have been considered as a non-conventional radiation probe in various applications. To utilize cosmic ray muons in engineering applications, two important quantities, trajectory and momentum, must be known. The muon trajectories are easily reconstructed using two-fold detector arrays with a high spatial resolution. However, precise measurement of muon momentum is difficult to be achieved without deploying large and expensive spectrometers such as solenoid magnets. Here, we propose a new method to estimate muon momentum using multi-layer pressurized gas Cherenkov radiators. This is accurate, portable, compact (< 1m$^3$), and easily coupled with existing muon detectors without the need of neither bulky magnetic nor time-of-flight spectrometers. The results show that not only our new muon spectrometer can measure muon momentum with a resolution of ±0.5 GeV/c in a momentum range of 0.1 to 10.0 GeV/c, but also we can reconstruct cosmic muon spectrum with high accuracy (~90%).

*Corresponding Author: Junghyun Bae
Email: bae43@purdue.edu

**Introduction**

Cosmic ray muons present a large part of background radiation[1] and they have recently been explored as a non-conventional radiation probe for monitoring or imaging the contents of dense and large objects, otherwise not feasible with conventional radiographic techniques[2]. It is noteworthy how the recent developments on cosmic ray muon applications have been successfully acknowledged in various engineering fields, including nuclear reactor and spent nuclear fuel cask imaging[3–6], homeland security[7–10], geotomography[11–15], and even archeology[16]. Despite the potential and success, the wide application of cosmic ray muons is limited by the naturally low muon intensity, approximately $10^4$ m$^{-2}$ min$^{-1}$ at sea level. Since it is not practical to deploy neither a large hadron collider nor an accelerator in the field to provide muon beams, it is important to measure muon momentum to maximize the utilizability of each cosmic ray muon[17,18]. However, it is still challenging to measure muon momentum in the field without resorting to large solenoid or toroidal magnets, Cherenkov ring imagers, or time-of-flight detectors[19–21]. Although recent efforts at CNL and INFN aim to infer momentum knowledge from multiple Coulomb scattering measurements[22,23], at present no portable spectrometer exists that can measure muon momentum in the field.

Here, we develop a new concept to measure muon momentum using multi-layer pressurized gas Cherenkov radiators. Because muons are charged particle, they can induce Cherenkov radiation in transparent media. Unlike solid or liquid Cherenkov radiators, the refractive index of gas radiators can be varied by changing gas pressure and temperature. We can then find the optimal muon threshold momentum levels by carefully selecting the gas pressure for each radiator. As a result, radiators will emit only when the threshold momentum is less than the actual muon momentum even though a muon passes through all radiators (Fig. 1). By measuring the Cherenkov signals in each radiator, we can then estimate the actual muon momentum (Fig. 2). The benefit of such an approach is compact, lightweight, and portable spectrometer that can be deployed in the field to improve existing or explore new engineering applications: cosmic ray muon tomography, geological studies, and cosmic radiation measurement in the International Space Station.

In this paper, we present the design concept, performance, and feasibility of our proposed muon spectrometer. This includes (i) theoretical background and (ii) principle of operation of the Cherenkov spectrometer and (iii) Geant4 modeling and simulations. The simulation results demonstrate that our spectrometer can measure muon momentum with high accuracy (~ 90%) for a wide muon momentum range (0.1 – 10.0 GeV/c) and a resolution of ±0.5 GeV/c which are sufficient for engineering applications. At the end, we demonstrate the benefit of muon momentum measurement in two real-world applications, (i) monitoring and (ii) imaging of special nuclear materials (SNMs) using muon multiple Coulomb scattering. The results show that SNM monitoring and imaging capability can be significantly improved when we integrate muon momentum knowledge into the existing technologies.

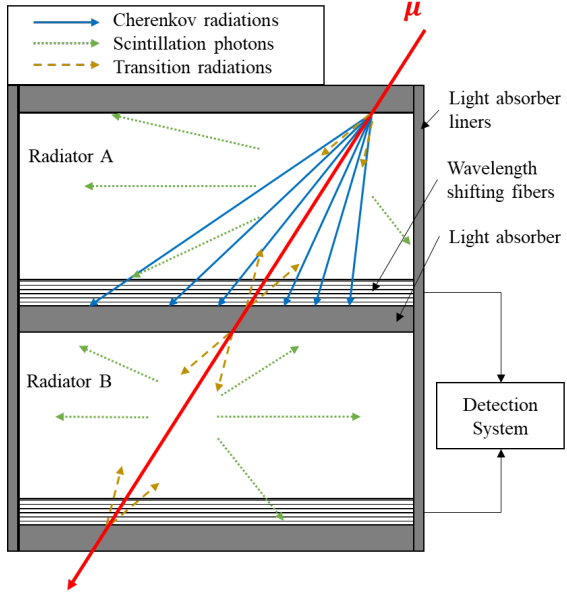 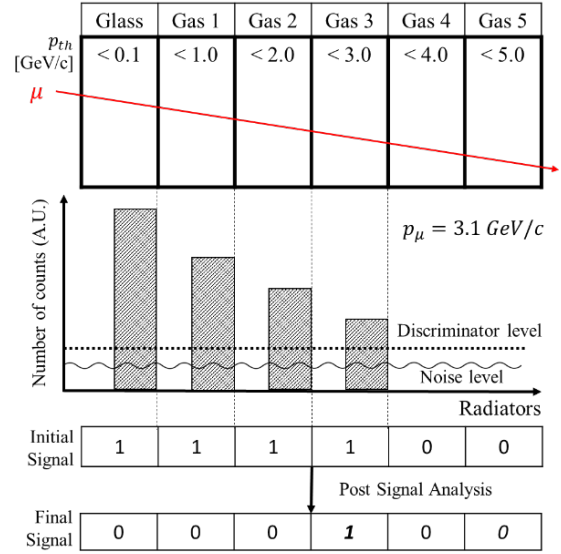

**Fig. 1 Characteristics of photon emission.** Cherenkov radiation, scintillation, and transition radiation can be emitted in radiators. Radiator A (top) emits Cherenkov photons since $p_\mu > p_{th}$ whereas radiator B (bottom) does not because $p_\mu < p_{th}$. However, both radiators emit scintillation and transition radiation regardless of actual muon momentum. Due to the forward-biased directional photon emission of Cherenkov radiation, scintillation photon signals can be efficiently discriminated.

**Fig. 2 Signal flow diagram.** The simplified schematic of signal flow to estimate muon momentum. In this example, only first four radiators emit Cherenkov radiation because the actual muon momentum is 3.1 GeV/c. Regardless of muon momentum, there always exists noise due to the scintillation, transition radiation, and electronic noise. Using a linear discriminator it correctly identifies the actual muon momentum with a resolution of ±0.5 GeV/c[24].

**RESULTS**

Various mono-energetic muons are simulated to evaluate the functionality of the muon spectrometer. The measured photon signals are classified into two categories, Cherenkov photons (signals) and others (noise). In the first simulation, optical photon emissions by the scintillation and transition radiation are excluded to focus on the Cherenkov photon emission by muons. Two mono-energetic muons, $p_\mu$ = 3.25 and 10.0 GeV/c, vertically enter the muon spectrometer and pass through all radiators (Fig. 3). When $p_\mu$ = 3.25 GeV/c, only the first four radiators emit conical shaped Cherenkov radiation because their threshold momenta are less than the simulated muon momentum. On the other hand, all radiators emit Cherenkov radiation when $p_\mu$ = 10.0 GeV/c because threshold momentum levels for all radiators are lower than 10.0 GeV/c. In the next simulation, scintillation and transition radiation photons are included in addition to Cherenkov radiation, however, all other parameters remain unchanged. More optical photons are observed in radiators due to the presence of scintillation and transition radiation. It is noted that scintillation photon emission can be efficiently differentiated from Cherenkov photons because of a) the characteristic photon emission direction ($\theta_c$ vs $4\pi$) and b) light flash duration (Fig. 4). There was no transition radiation observed in Fig. 3 because it is a rare event for a few GeV muons and for the small number of boundary systems.

It is noteworthy that some additional Cherenkov photons ($N_{ch} \neq 0$ when $p_\mu < p_{th}$) are observed in Fig. 3 and they are also recorded (Fig. 5). It is because of either Compton scattering or Cherenkov photon emissions by secondary particles (mainly electrons) which can be produced by muon decays and muon to electron conversions[25].

$$\mu^- \rightarrow e^- \bar{\nu}_e \nu_\mu \qquad (1)$$

$$\mu^+ \rightarrow e^+ \nu_e \bar{\nu}_\mu \qquad (2)$$

$$\mu N_{Al} \rightarrow e N_{Al} \qquad (3)$$

Equation (1) and (2) represent radiative decays of $\mu\pm$ with a mean lifetime of $\tau_\mu \cong 2.2$ $\mu$sec. Radiative muon decays are the primary source of secondary charged particles. Equation (3) represents the muon to electron conversion. Muons can be captured by the *1s* orbital of aluminum and mono-energetic electron emission follows [26]. Although the muon capture cross-section is insignificant, it is worth to be considered because the photon absorbers are made of aluminum foils.

The results of total number of optical and Cherenkov photons as a function of muon energy are shown in Fig. 5. The rapid increases of total number of emitted photons at the muon threshold momentum levels (vertical dashed lines) are noted and they are results of the Cherenkov radiation by muons. Then, it gradually increases as muon energy increases. It demonstrates that our proposed spectrometer is feasible because muon energies are clearly identified by analyzing optical photon signals from radiators.

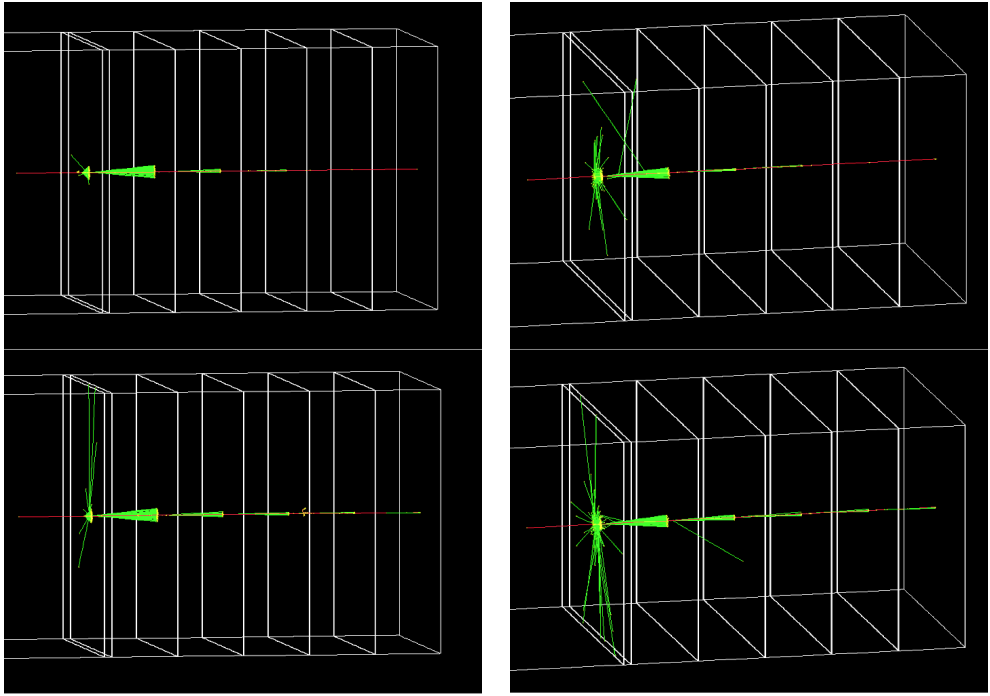

**Fig. 3 Geant4 simulation results:** (i) Cherenkov radiation only (left column), (ii) Cherenkov radiation, scintillation, and transition radiation (right column) when muon momenta are 3.25 (top row) and 10.0 GeV/c (bottom row). Green and red lines represent optical photons and negative muons, respectively.

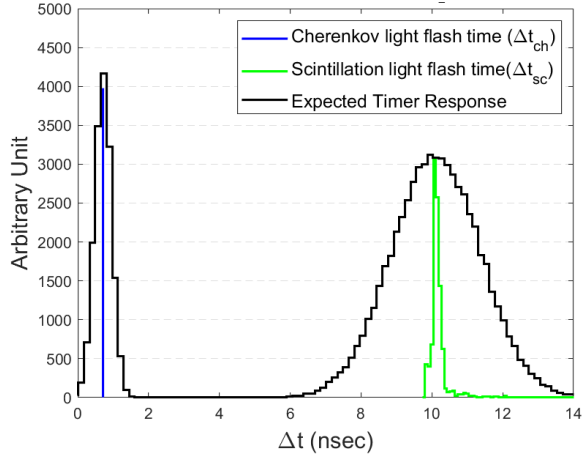 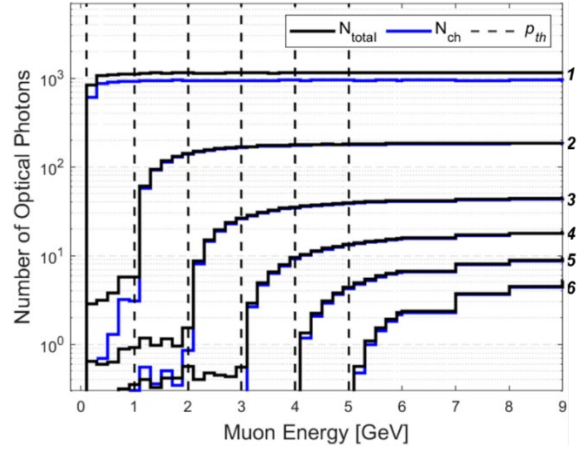

**Fig. 4 Cherenkov and scintillation light flash time.** Cherenkov light flashing time has a fast pulse ($< 1\times10^{-9}$ sec) because it is the result of instant physical disorder caused by the incident muon whereas scintillation light flash time has a slow distribution with a tail due to the decay constant ($\gg 1\times10^{-9}$ sec) because it is a fluorescence process of scintillation.

**Fig. 5 Expected number of optical photon emission.** The rapid increments are observed at the threshold momentum levels (vertical dashed lines) because of Cherenkov radiation emission above the threshold momentum levels. Then, it gradually increases as muon energy increases. The italic numbers (1 to 6) on the right represent the radiator IDs.

**Performance Evaluation**

We evaluate the performance of our muon spectrometer by reconstructing a cosmic ray muon spectrum. The experimental results and mathematical expressions for cosmic ray muon spectrum at sea level are found in references [27–29]. We used the "Muon_generator_v3" [30] which is developed to simulate cosmic ray muon spectra based on validate semi-analytical models [31,32]. To quantify the muon spectrometer measurement accuracy, we introduced a classification rate (CR) which is a probability that the system correctly identifies the actual muon momentum. Reconstructed cosmic muon spectra (< 10 GeV/c) using $10^4$ muon samples with (i) perfect and (ii) practical (noise is included) muon spectrometers are shown in Fig. 6. The computed CRs for the practical muon spectrometer are also presented.

The next simulation results are the computed CRs using $10^4$ mono-energy muons as a function of muon momentum with various discriminator levels, 0, 1, and 2 (Fig. 7). Simple linear discriminators were used in the final stage of signal process to cut off predictable noise from total signals. When $p_{th} < 1.0$ GeV/c, overall CR is less than 60% because of high scintillation photon intensity in the glass radiator. Hence, a CR increases as using higher discrimination levels in this level. When $1.0 < p_{th} < 4.0$ GeV/c, all gas radiators show high CRs because Cherenkov and scintillation signals are clearly differentiated using a linear discriminator. However, when $4.0 < p_{th} < 6.0$ GeV/c, CRs begin to decrease since Cherenkov photon yields are too low due to the rarefied $CO_2$ (~ a half atmosphere pressure) gas for high muon threshold momentum. When $p_{th} > 6.0$ GeV/c, on the other hand, CR rebounds because Cherenkov photon yields gradually increase as muon momentum increases. However, it is not efficient to use any linear discriminator in high $p_{th}$ (> 5.0 GeV/c). In addition, it is noted that there are repeated sudden drops of CR in Fig. 7. This is because $p_\mu$ near $p_{th}$ boundaries, 0.1, 1.0, … 5.0 GeV/c, tends to have a high false classification rate.

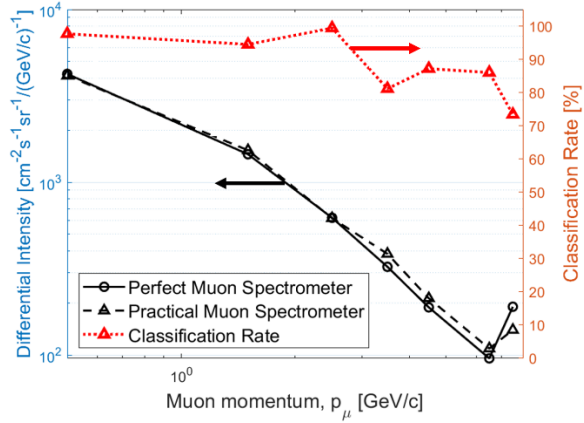 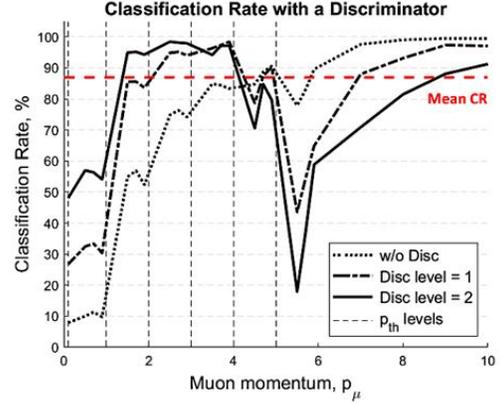

**Fig. 6 Reconstructed comic ray muon spectra.** Reconstructed cosmic ray muon spectra are reconstructed using $10^4$ muon samples with (i) perfect and (ii) practical muon spectrometers. Differential muon intensity increases and classification rate decreases at > 5.0 GeV/c because all muons that have a momentum greater than 5.0 GeV/c are categorized in the largest $p_{th}$ level. The computed classification rates for measurement classes are also indicated.

**Fig. 7 Classification rate as a function of muon momentum.** Using $10^4$ mono-energy muons, the classification rates are plotted as a function of muon momentum with various discriminator levels, 0, 1, and 2, from 0.1 to 10.0 GeV/c. Repeated sudden drops of CR are observed because muon momentum near $p_{th}$ boundaries tends to have a high false classification rate.

**Benchmark**

*Separation and Identification of SNMs*

One potential application of our proposed muon spectrometer is to monitor hidden special nuclear materials. Cosmic ray muons have been considered as an unconventional radiation probe for monitoring of SNMs due to their high-penetrative nature and their charge. Muons are not easily absorbed by SNMs but undergo multiple Coulomb scattering (MCS) and they are deflected with characteristic angles depending on the Z number of materials and momentum. The estimated muon scattering distribution using the MCS approximation is known to follow Gaussian distribution with a zero mean and it is given by [33]:

$$f(\theta|0, \sigma_\theta^2) = \frac{1}{\sqrt{2\pi}\sigma_\theta} \exp\left(-\frac{1}{2}\frac{\theta^2}{\sigma_\theta^2}\right) \quad (4)$$

$$\sigma_\theta = \frac{13.6 \, MeV}{\beta_\mu p_\mu c} \sqrt{\frac{X}{X_0}} \left[1 + 0.088 \log\left(\frac{X}{X_0}\right)\right] \quad (5)$$

where $\theta$ is the muon scattering angle, $\sigma_\theta$ is the standard deviation of scattering angle, $X$ is the thickness of the scattering medium, and $X_0$ is the radiation length.

To investigate the effect of including muon momentum in SNM monitoring, high enriched uranium (HEU), low enriched uranium (LEU), plutonium (Pu), and lead (Pb) shielding were studied using Monte Carlo simulations. We used $10^3$ muons (approximately translate to a few minutes of scanning time) to analyze their scattering angle variance distributions in order to identify SNMs and separate non-SNM from them. The results for SNM separation and identification with and without muon momentum knowledge are shown in Fig. 8. By measuring muon momentum, Pb is clearly separated from SNMs and identification of each type of SNM is also possible even when they are surrounded by 30 cm of lead shielding. This last scenario was almost impossible without muon momentum knowledge.

*Muon Scattering Tomography*

Our proposed muon spectrometer can be incorporated with a muon scattering tomography system which reconstructs images of hidden high-density materials using the point-of-closest approach (PoCA) algorithm [34]. As shown in (5), the expected muon scattering angle distribution is determined by two variables, material property and muon momentum. In other words, materials can be identified by measuring muon scattering angle distribution and momentum. Using a momentum integrated PoCA (mPoCA) algorithm which includes the muon momentum term, we successfully improved the image resolution as shown in Fig. 9. The experimental configuration includes cylindrical tungsten (radius = 5.5 cm, height = 5.7 cm) at the center and two stainless beams on the sides. This example was firstly introduced by K. Borozdin et al[35]. Without muon momentum measurement, many incorrect voxels are found between tungsten and steel beams. However, the resolution of reconstructed image is significantly improved by coupling muon momentum knowledge with PoCA algorithm.

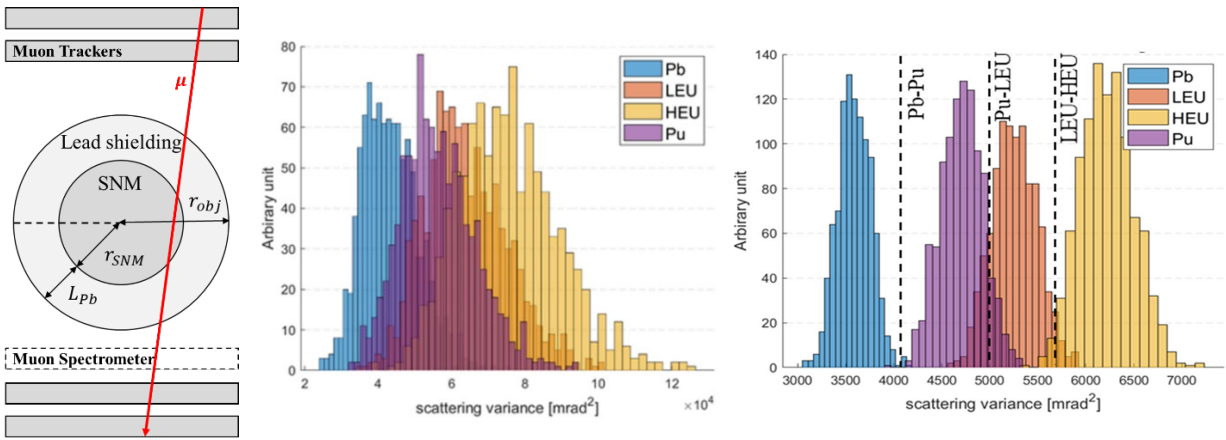

**Fig. 8 Muon scattering angle variance distributions.** Schematic drawings of the SNM monitoring using cosmic ray muons (left). Scattering angle variance distribution for HEU, LEU, Pu, and Pb using $10^3$ muon samples (it can translate to ~min of scanning time) without (center) and with (right) muon momentum knowledge. By measuring muon momentum, we can separate lead from SNMs and identify each type of SNM even when they are surrounded by thick lead shielding ($L_{Pb}$ = 30 cm)[18].

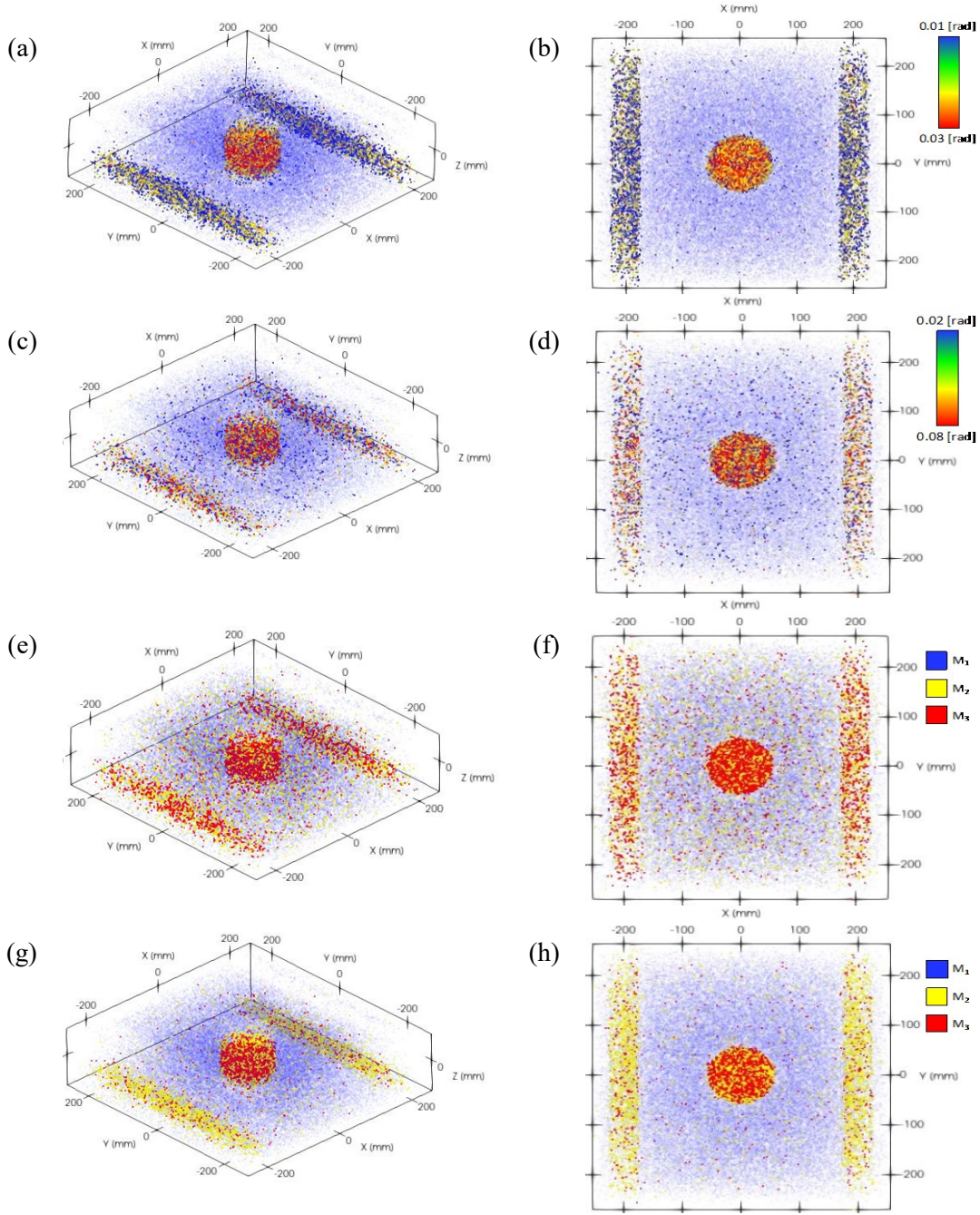

**Fig. 9 Reconstructed images using a modified PoCA.** The results of image reconstruction using muon tomography (a cylindrical tungsten and two steel beams) for four scenarios using original PoCA (a – d) and mPoCA (e – h): $10^5$ mono-energy muons, 3 GeV (a, b), (ii) $10^5$ muons with cosmic muon spectrum (c, d), (iii) $5\times10^4$ muons (e, f), and (iv) $10^5$ muons (g, h) with cosmic muon spectrum. It is noted that continuous and discrete color maps are used for PoCA and mPoCA, respectively. Types of materials are indistinguishable when muon momentum knowledge is not considered.

**DISCUSSION AND CONCLUSION**

In this paper, we presented a new concept to measure muon momentum using multi-layer pressurized gas Cherenkov radiators. Since our proposed muon spectrometer relies on pressurized gas Cherenkov radiators, we can eliminate the need for bulky and expensive magnets to measure muon momentum. Six sequential threshold momentum levels were set using a solid radiator ($SiO_2$) and pressurized $CO_2$ gases, $p_{th}$ = 0.1, 1.0, 2.0, 3.0, 4.0, and 5.0 GeV/c. Each radiator can emit Cherenkov radiation when a muon passes through it if the actual muon momentum exceeds the threshold muon momentum of this radiator. On the other hand, radiators will emit scintillation photons and transition radiation even though the actual muon momentum is lower than the threshold momentum.

We performed detailed Geant4 simulations to assess the performance and demonstrate the functionality of the proposed muon spectrometer by investigating the Cherenkov, scintillation, and transition radiation photon emission in the radiators. Using our proposed muon spectrometer, we were able to measure muon momentum with a resolution of ±0.5 GeV/c within the momentum ranges of 0.1 to 10.0 GeV/c. This is sufficient for most engineering applications. Moreover, a cosmic ray muon spectrum was successfully reconstructed with a high accuracy (~90%).

To further demonstrate the potential of measuring muon momentum in special nuclear material monitoring and muon scattering tomography applications, we compared the results of both applications with and without muon momentum measurement. In the SNM monitoring application, we showed that we are not only able to separate lead from SNMs but also identify each type of SNMs (HEU, LEU, and Pu) which was not possible without measuring muon momentum. In the muon scattering tomography, we developed a momentum integrated PoCA algorithm (mPoCA) in order to couple muon momentum with the original PoCA algorithm. When we reconstructed using the mPoCA algorithm, imaging resolution was significantly improved.

**METHODS**

**Cherenkov muon spectrometer**

A threshold muon momentum is required to induce Cherenkov radiation in a given radiator and it is derived as follows [36]:

$$p_{th}c = \frac{m_\mu c^2}{\sqrt{n^2 - 1}} \quad (6)$$

where $p_{th}$ is the threshold muon momentum, $n$ is the refractive index of the medium, and $m_\mu c^2$ is the muon rest mass energy. The threshold momentum for muons is determined by the refractive index of the radiator. In gas radiators, the refractive index depends on both pressure and temperature and it is approximated by the Lorentz-Lorenz equation:

$$n \approx \sqrt{1 + \frac{3A_m p}{RT}} \quad (7)$$

where $A_m$ is the molar refractivity, $p$ is the gas pressure, $T$ is the absolute temperature, and $R$ is the universal gas constant. Therefore, the threshold muon momentum is given by:

$$p_{th} = m_\mu c \sqrt{\frac{R}{3A_m}\frac{T}{p}} \tag{8}$$

Equation (8) shows that muon threshold momentum can be varied by changing the pressure or temperature without replacing radiator materials. For the gas selection, we considered, $C_4F_{10}$, $C_3F_8$, R-12, and $CO_2$. $C_4F_{10}$ and $CO_2$ are used in the Thomas Jefferson National Laboratory as gas Cherenkov radiators [37]. R-12 gas refrigerant was used for gas radiators in the past, however, the usage is limited due to its high Ozone Depletion Potential (ODP). Hence, a refrigerant, R-1234yf ($C_3H_2F_4$), becomes a promising alternative gas radiator to replace R-12 ($CCl_2F_2$) [38]. $C_3F_8$ is the alternative gas radiator for $C_4F_{10}$ due to the better stability at the high pressure. Gas pressure cannot exceed 3 atm without condensation using the $C_4F_{10}$ at room temperature.

**Prototype Design**

The underlying principle of the proposed muon spectrometer is illustrated in Fig. 1. It shows characteristics of photon emissions by Cherenkov, scintillation, and transition radiation. All surfaces of gas radiator tanks are surrounded by strong photon absorbers so that emitted photons are isolated within radiators. We chose $CO_2$ gas because it can be pressurized up to approximately 5.7 MPa without condensation although the refractive index of $CO_2$ gas is relatively lower than other gas radiators. Five sequential threshold momentum levels are obtained using pressurized $CO_2$ gas radiators, $p_{th}$ = 1.0, 2.0, …, 5.0 GeV/c, however, we inevitably had to use a solid radiator ($SiO_2$) for the lowest threshold momentum level, $p_{th}$ = 0.1 GeV/c, because it is impossible to reach that threshold momentum level with pressurized $CO_2$ at room temperature.

**Geant4 Simulations**

To demonstrate the functionality of proposed muon spectrometer, a high-fidelity stochastic muon transport simulation using Geant4 was performed. The overall length of the muon spectrometer is 51.7 cm and the surface area is 20 × 20 $cm^2$. It consists of one glass radiator (1 cm), five $CO_2$ gas radiators (10 cm), and thin photon absorber foils ($10^{-1}$ cm). The overall weight of muon spectrometer is less than 10 kg because it mostly consists of gas radiators but structural frames (stainless steel) and optical sensors weigh approximately 8 kg. For glass radiator, a standard $SiO_2$ with refractive index of 1.45 is chosen, and for gas radiators, sequentially pressurized $CO_2$ gases are used to provide desirable refractive indices and muon threshold momenta. Some $CO_2$ gas radiators are technically depressurized because Cherenkov threshold momentum at atmospheric pressure is approximately 3.5 GeV/c. Therefore, some are depressurized to achieve $p_{th}$ = 4 and 5 GeV/c. Any optical photon events are terminated once they arrive either at the photon absorbers or outside of "world" volume during the simulations. The overview of field muon spectrometer configuration is shown in Fig. 10. All major physics are included in the Geant4 reference physics list, QGSP_BERT [39].

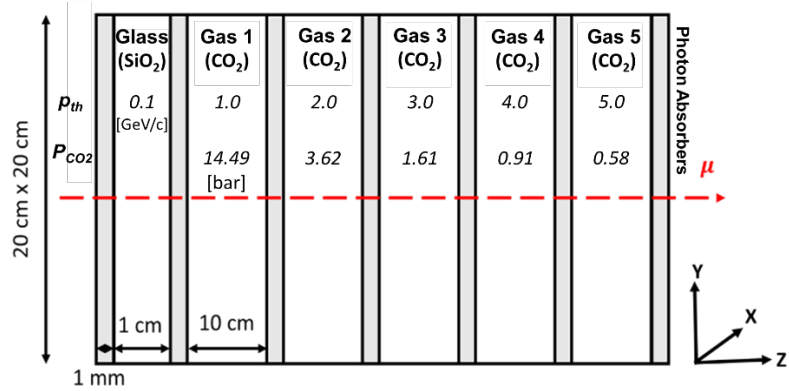

**Fig. 10 Overview of the fieldable muon spectrometer.** The schematic of proposed muon spectrometer. It shows dimensions and materials of components. In addition, threshold momentum and $CO_2$ gas pressure of each radiator are also presented. Note: The figure is not proportional to the actual size.

**Signal Analysis**

Optical sensors and photon absorbers are installed on one side of each radiator. Photon absorbers prevent photons from escaping and help efficiently discriminate scintillation photon signals given that scintillation photons are emitted in all directions whereas Cherenkov photons are emitted with a forward-biased direction. Fig. 2 shows an example of signal flows when a single muon ($p_\mu$ = 3.1 GeV/c) passes through all radiators. First four radiators emit Cherenkov radiation ("triggered") because their threshold momentum levels are lower than the incident muon momentum, 3.1 GeV/c. In this example, the system records '1' for triggered radiators and '0' for intact radiators. Due to the presence of scintillation and transition radiation, noise signals always exist. Therefore, a use of discriminator is suggested to efficiently remove predicted noise from signals. After the signal process, a final signal successfully identifies the actual muon momentum with a resolution of ±0.5 GeV/c, even though the muon momentum is very close to the threshold momentum level.

**Acknowledgment**

This research is being performed using funding from the Purdue Research Foundation.

**Author contributions**

J. B. carried out the analytical derivations, numerical computations, and paper writing. S. C. conceived the proposed scheme and analyzed the results of numerical simulations. All authors reviewed the paper.

**Competing interests**

The authors declare no competing interests.

**Additional Information**

Muon generator_v3 is available online: https://www.mathworks.com/matlabcentral/fileexchange/51203-muon-generator.

**Correspondence** and requests for materials should be addressed to J. Bae.